\documentstyle[12pt]{article}
\begin{document}

\begin{center}
\Large{\bf COMMENTS ON 
           BIOLOGICAL EFFECTS OF STELLAR COLLAPSE NEUTRINOS} 
\end{center}
\vskip 3 cm

\begin{center}
{\bf Samar Abbas $^{* \#}$ and Afsar Abbas $^\#$}

\vskip 5 mm

$\#$ Institute of Physics, Bhubaneswar-751005, India

$*$ Physics Dept., Utkal University, Bhubaneswar-751004, India

(e-mail : abbas@iopb.ernet.in, afsar@iopb.ernet.in)
\end{center}

\vskip 30 mm

\begin{centerline}
{\bf Abstract }
\end{centerline}
\vskip 5 mm

Extraterrestrial processes like neutrinos from collapsing 
stars, cosmic rays from supernovae and cosmic rays from neutron 
star mergers etc. have recently been proposed as models to explain
the periodic mass extinctions such as that which wiped out the
dinosaurs at the K-T boundary. Here we show that these models fail
to give any reasonable explanation of several empirically 
established facts related to these mass extinctions.

\newpage

Collar in a recent paper [1] has attempted to explain the 
periodicity of mass extinctions as arising from stellar collapse 
neutrinos. In the final stages of collapse, a massive star 
radiates most of its binding energy in the form of MeV neutrinos. 
He calculated the frequency of such
stellar collapses in the vicinity of the Solar Sytem
which was then suggested to correspond to 
the observed periodicity of 
palaeontological mass extinctions.

The MeV neutrinos interact with matter through 
elastic collisions via
the neutral currents. This causes radiation damage in organic 
tissues.
`...every $ \sim $ 100 Myr a collapse 
within 6.3 pc of the Earth will 
produce $ \sim 1.9*10^4 $ recoils / kg in all living tissue' (p.1000).
This would lead to irreparable DNA damage.
`This is a $ \sim $ 4 (12) malignant foci/kg tissue 
occurring every $ \sim $ 30 (100) Myr,
an insult that would be severe enough to kill a vast percentage
of large animals with a frequency comparable to that of most major 
extinctions' (p.1001).
 This is the main conclusion of the paper [1].

In this comment we would like to point out that 
the scenario envisioned
above cannot possibly explain the five major mass
extinctions or the 30 to 100 Myr periodicity of mass extinctions
in the last 600 Myr. There are several fundamental 
and empirically established aspects of the
mass extinctions which the paper [1] fails to explain and hence may
be ruled out.

The Cretaceous-Tertiary, or K-T
mass extinction that took place 65 Myr ago
is often attributed to either the impact of an asteroid/comet [2]
or to massive volcanism [3]. Both these models assume that dust and
chemicals lofted into the atmosphere caused climatic changes in the 
ocean and atmosphere, thereby playing havoc with the prevailing
ecosystem and leading to the eradication of large numbers of plant
and animal families.
The debate between the two hypotheses still continues [3-6]. 
The iridium anomaly 
at the K-T boundary was the motivating concept for the impact
hypothesis but is also explainable in terms of the volcanic 
hypothesis.

The iridium anomaly is an empirically well
established aspect of the K-T
extinction [7] and has also been 
reported for some of the other 
significant extinctions. Collar's model [1] fails to
give any reasonable explanation of these iridium anomalies.

The large-scale flood basalt volcanism that occurred during 
several significant mass extinctions,
such as the Deccan and Brito-Arctic flood basalts which
were deposited at the K-T boundary, has been linked to these
mass extinctions [3,6]. No possible
understanding of flood basalt volcanism arises in the framework
of the paper under discussion [1].

The existence of microspherules and shock deformation features
associated with the K-T mass extinction cannot be explained by this
model [1].

The regression of the seas associated with several mass extinctions
[3,7] cannot be explained by Collar's model [1].

Moreover the presumed MeV neutrino bursts leading to malignant
tissues would occur during a very short time ($ < $ 10 sec) 
[1,p.999].
So as per this model the mass extinctions would occur almost 
instantaneously. However this is inconsistent with the fact
that extinctions are believed to have occurred over tens to
hundreds of thousands of years [3,4,6,7].

As per this model [1] large animals die `while smaller forms
may be spared by not enough of their members being directly
killed for a non-viable minimal population to be reached'
[1,p.1001]. This is true in this model for the 
land as well as for the 
sea [p.1002]. While this has been established to be 
true for the land mass extinctions, surprisingly 
the opposite holds 
true for the sea mass extinctions [8].
It has been found that large predators survived in the sea while the
smaller ones perished [8]. This fact goes against the model under
review [1].

Hence it is obvious that Collar's model [1] fails to
give any reasonable explanation for several fundamental
empirically established features of the five major and the other
periodic mass extinctions and is hence not viable.

For the sake of completeness one may mention that the other
related supernova extinction hypotheses [9] cannot possibly explain 
the regression of the sea and the eruption of the Siberain flood
basalts associated with the Permo-Triassic mass extinction [10].
The recently proposed concept of mass extinctions as arising from
neutron star mergers [11] suffers from the same shortcomings
as discussed above in the context of Collar's work.

\newpage
{\sc REFERENCES}

\vskip 10 mm

1. J.I.Collar, Phys.Rev.Lett., {\bf 76 } (1996) 999

2. L.W. Alvarez, W. Alvarez, F. Asaro and H.V. Michel, 
   
   Science, {\bf 208 } (1980) 1095

3. C.B. Officer, A. Hallam, C.L. Drake and J.D. Devine, 
  
   Nature {\bf 326 } (1987) 143

4. P. Hut, W. Alvarez, W.P. Elder, T. Hansen, E.G.Kauffman,G.Keller,
   
   E.M. Shoemaker and P.R. Weissman , Nature {\bf 329 } (1987) 118

5. W. Alvarez and F. Asaro, Sc. Am. Oct.1990, p.78

6. V.E. Courtillot, Sc. Am. Oct.1990, p.85

   M.R. Rampino and R.B. Stothers, Science {\bf 241} (1988) 663

7. N. Bhandari, Current Scince (India) {\bf 61 } (1991) 97

8. R. Bakker, Discover (Sept.1993), p.78

9. J. Ellis and D. Schramm, Proc. Natl. Acad. Sci. USA 
   {\bf 92 } (1995) 235

10. D.H.Erwin, Nature {\bf 367 } (1994) 231

11. A. Dar, A. Laor and N.J. Shaviv, Haifa Preprint 1996,

    astro-ph/9607160

\end{document}